\documentclass[11pt]{article}

\usepackage{cite}

\usepackage[round]{natbib}
\usepackage{amsmath}
\usepackage{amssymb}

\usepackage{graphicx}

\usepackage{color}

\usepackage{setspace}

\usepackage[labelfont=bf,labelsep=period,justification=raggedright]{caption}

\makeatletter
\renewcommand{\@biblabel}[1]{\quad#1.}
\makeatother

\date{}

\usepackage{subfigure}
\usepackage{multirow}
\usepackage{pdflscape}
\usepackage{rotating}

\usepackage{caption}

\begin{document}

\begin{titlepage}

\begin{center}
{\LARGE Brain networks reveal the effects of antipsychotic drugs on schizophrenia patients and controls} \\*[1cm]
{\large Emma K. Towlson$^{1,2}$, Petra E. V\'{e}rtes$^3$, Ulrich M\"{u}ller$^{4,5}$, and Sebastian E. Ahnert$^{6,7}$}\\*[0.25cm]
\end{center}

\begin{enumerate}

\item Center for Complex Network Research and Department of Physics, Northeastern University, Boston, Massachusetts 02115, USA
\item Media Laboratory, Massachusetts Institute of Technology, Cambridge, MA 02139, USA
\item Department of Psychiatry, Behavioural and Clinical Neuroscience Institute, University of Cambridge, Cambridge CB2 0SZ, UK
\item Department of Psychology, University of Cambridge, Downing Street, Cambridge, CB2 3EB, UK 
\item Barnet Enfield Haringey NNHS Mental Health Trust, Block B2, St. Ann's Hospital, St Ann's Rd, London N15 3TH, UK
\item Theory of Condensed Matter Group, Department of Physics, Cavendish
Laboratory, University of Cambridge, JJ Thomson Avenue, Cambridge, CB3 0HE, UK
\item Sainsbury Laboratory, Cambridge University, Bateman Street, Cambridge, CB2 1LR, UK

\end{enumerate}

\paragraph*{Correspondence:}  Emma Towlson. Center for Complex Network Research and Department of Physics, Northeastern University, Boston, Massachusetts 02115, USA. \\
Email: ektowlson@gmail.com

\paragraph*{Acknowledgements:}
EKT was supported by an Engineering \& Physical Sciences Research Council (UK) PhD studentship, and is now supported by NSF award number 1734821. PEV was supported by the Medical Research Council (MR/K020706/1), and is now a Fellow of MQ: Transforming Mental Health, grant number MQF17\_24. The Behavioural \& Clinical Neuroscience Institute is supported by the Medical Research Council (UK) and the Wellcome Trust. SEA was supported by a Royal Society University Research Fellowship, and is currently supported by a Gatsby Career Development Fellowship. Data collection was supported by a grant from Bristol Myers Squibb to the late Robert Kerwin at King's College London. We thank Ameera Patel for discussions about pre-processing the raw fMRI data.

\paragraph*{Authors' contributions:} EKT and SEA conceived of the analysis. UM designed the clinical trials, recruited the subjects, and gathered the data.  EKT did the analysis, SEA oversaw it, and EKT, SEA, and PEV interpreted the results. EKT, PEV, and SEA wrote the manuscript.

\paragraph*{Conflict of interest:} UM has received honoraria for advisory board participation, consultancy and educational talks, travel expenses and/or support for educational conferences from Astra Zeneca, Bristol Myers Squibb, Eli Lily, Heptares, Lundbeck, Flynn Pharma/Medice, Janssen-Cilag, Shire and/or Sunovion.

\end{titlepage}

\newpage

\begin{abstract}

The study of brain networks, including derived from functional neuroimaging data, attracts broad interest and represents a rapidly growing interdisciplinary field. Comparing networks of healthy volunteers with those of patients can potentially offer new, quantitative diagnostic methods, and a framework for better understanding brain and mind disorders. We explore resting state fMRI data through network measures, and demonstrate that not only is there a distinctive network architecture in the healthy brain that is disrupted in schizophrenia, but also that both networks respond to medication. We construct networks representing 15 healthy individuals and 12 schizophrenia patients (males and females), all of whom are administered three drug treatments: (i) a placebo; and two antipsychotic medications (ii) aripiprazole and; (iii) sulpiride. We first reproduce the established finding that brain networks of schizophrenia patients exhibit increased efficiency and reduced clustering compared to controls. Our data then reveals that the antipsychotic medications mitigate this effect, shifting the metrics towards those observed in healthy volunteers, with a marked difference in efficacy between the two drugs. Additionally, we find that aripiprazole considerably alters the network statistics of healthy controls. Using a test of cognitive ability, we establish that aripiprazole also adversely affects their performance. This provides evidence that changes to macroscopic brain network architecture result in measurable behavioural differences. This is the first time different medications have been assessed in this way. Our results lay the groundwork for an objective methodology with which to calculate and compare the efficacy of different treatments of mind and brain disorders.

\end{abstract}

\section*{Introduction}

In recent years neuroimaging data and graph theory have allowed for the description of the topological properties of large-scale brain networks \citep{bullmore2013, bullmore2009, vandenheuvel2010}. Disorders of the brain have long been thought to be due to abnormal connectivity patterns, and these networks allow for a quantitative measure of this disruption \citep{rubinov2013b, bullmore2012}. Schizophrenia is a debilitating psychiatric condition with a range of symptoms including auditory and visual hallucinations, delusions, disorganised thinking, and cognitive impairment. Various network-based studies have associated schizophrenia with a subtle randomisation of connections \citep{fornito2012,alexanderbloch2013, vandenheuvel2013, lynall2010, rubinov2013, yu2011}. Antipsychotic medications are employed to treat symptoms with varying degrees of success and side-effect. Sulpiride is a selective dopamine antagonist most commonly used in Europe and Japan for schizophrenia treatment. Aripiprazole is an atypical third generation antipsychotic introduced for the treatment of schizophrenia in the USA in 2002 and Europe in 2004 \citep{bolstad2015}. It acts as a dopamine receptor partial agonist, whereas typical antipsychotics used to combat the symptoms of schizophrenia are pure dopamine antagonists. Partial agonists have long been of interest \citep{lieberman2004} in order to avoid the extrapyramidal and endocrine side effects caused by typical antipsychotics.

To date, very few studies have been conducted to understand if and how medication alters an individual's brain network \citep{achard2007,hadley2016,flanagan2018}. We hypothesised that an effective medication would act to make the functional brain networks of patients more similar to those of healthy volunteers. We set out to test this in the context of three drug treatments: (i) placebo; (ii) aripiprazole, and  (iii) sulpiride. We used resting state fMRI data to analyse the functional connectivity, and a working memory task to assess the cognitive abilities, of 15 healthy volunteers and 12 patients with chronic schizophrenia.

Our results show that schizophrenia patients and healthy controls exhibit different network topologies, in agreement with the existing literature \citep{alexanderbloch2013, lynall2010}. Further, the drug treatments alter the topology of the brain network in a measurable way, both in healthy individuals and patients. In the brain networks of patients we found evidence that the drugs lead to network topologies that are closer to those of healthy individuals. This correlates with improved cognitive performance. In healthy individuals, treatment with aripiprazole leads to a significantly altered network as well as lower cognitive performance.

\newpage

\section*{Materials and Methods}

\subsection*{Experimental Design and Statistical Analysis}

\subsubsection*{Sample}

Twelve people with chronic schizophrenia and 15 healthy, nonpsychotic volunteers were recruited for participation in this study. The patients were diagnosed according to standard operational criteria in the DSM-IV \citep{DSMIV2000} and were clinically stable during their involvement (i.e. exhibiting low symptom ratings and undergoing no change of medication in the preceding four weeks). All were receiving antipsychotic drugs, and four were receiving additional psychotropic medication, but were not treated with their usual medication on the days of scanning to avoid effects on the fMRI data. Healthy volunteers were selected to match the patient group in terms of age, gender, premorbid IQ, years of education and handedness, and screened for major psychiatric disorders using the Mini International Neuropsychiatric Interview \citep{sheehan1998}. All subjects provided informed consent in writing and the protocol was approved by the Addenbrooke's NHS Trust Local Research Ethics Committee.

Every subject attended three scanning sessions, each one to two weeks apart, for collection of functional MRI data and completion of working memory tests (see below). At each visit they were administered one of three drug treatments: (i) an oral placebo 180 and 90 mins before scanning; (ii) oral aripiprazole 15mg 180 mins before scanning and oral placebo 90 mins before; (iii) oral placebo 180 mins before scanning and oral sulpiride 400 mg 90 mins before. We used a double dummy design with dosing of aripiprazole 180 mins and sulpiride 90 mins before the start of fMRI scanning. At both time points (-180 mins and -90 mins) we co-administered 10mg of Domperidone to minimise side effects (nausea). Both aripiprazole and sulpiride are antipsychotic medications designed to alleviate the symptoms of schizophrenia. The order of drug administration and the playlists of the working memory paradigm were counterbalanced across one group and repeated for the other.

\subsubsection*{Working memory tests}

At each session, the subjects were required to complete an ``N-back" task to assess their verbal working memory \citep{baddeley2003, koshino2005}. The task demanded that subjects maintain a series of visually presented letters in their working memory such that each stimulus could be compared to the letter presented N letters earlier in the series (i.e. N-back) - see Figure \ref{fig:fig2}(a). For example, if the sequence of letters was F-B-A-C, the subject could be expected to indicate on presentation of the last letter in the series that €œB€ was presented two letters previously (2-back). Difficulty was manipulated to 4 levels (0-back to 3-back) by varying the number of letters back in the series that the subject had to compare to the current letter. All subjects completed a practice version and three playlists matched for difficulty and distraction, which were allocated across sessions such that each subject would complete each playlist once across their three visits (for placebo, aripiprazole, and sulpiride). An individual's performance at this cognitive task was assessed by recording their hit rate, defined as the proportion of times they were able to successfully present a correct answer. In each session, there were 10 correct targets for each N-back level. Data for the performance of patients 11 and 12 administered sulpiride were missing so analyses were carried out without them.

\subsubsection*{Acquisition and preprocessing of fMRI data}

A General Electric (GE) Signa system scanner operating at 1.5 T at the BUPA Lea Hospital (Cambridge, UK)) was used to acquire functional MRI data over 17 min 12s, during which time subjects were asked to lie quietly with their eyes closed. In each session, 516 gradient-echo T2*-weighted echo planar images depicting blood oxygenation level-dependent (BOLD) contrast were generated from 16 noncontiguous near-axial planes: repetition time $=$ 2 s, echo time $=$ 40 ms, flip angle $=  70 \,^{\circ}$, voxel size $=$  3.05 $\times$ 3.05 $\times$ 7.00 mm, section skip $=$  0.7 mm, matrix size $=$ 64 $\times$ 64, field of view (FOV) $=$  240 $\times$ 240 $\times$ 123 mm. 4 volumes were discarded to allow for T1 equilibration effects, leaving 512 volumes per dataset \citep{lynall2010}.

Control 2 was missing an anatomical image so was discarded from the study, and patient 11 was missing data for the aripiprazole treatment. Each dataset was analysed for effects of head motion within the scanner \citep{power2012, vandijk2012, satterthwaite2012}, resulting in the further rejection of patient 3 on aripiprazole and sulpiride, control 10 on placebo and sulpiride, control 8 on sulpiride, and patient 5 on placebo, all of which were deemed to have too many motion related artefacts to be reliable. The remaining datasets were corrected for motion through realignment and wavelet despiking \citep{suckling2006,patel2014}. We used a 12 parameter affine transformation to register the data to MNI stereotactic standard space and a 6mm Gaussian kernal for spatial smoothing. Finally, the voxel timeseries were high- and low-pass filtered with cutoff frequencies of $\approx$ 0.01 Hz and $\approx$ 0.08 Hz respectively.

\subsubsection*{Kolmogorov-Smirnov test}

The two-sample \emph{Kolmogorov-Smirnov test} is a non-parametric test to compare two sets of data. The Kolmogorov-Smirnov statistic is a measure of the distance between the empirical distribution functions of the two samples and is calculated under the null hypothesis that both samples are taken from the same continuous probability distribution. The statistic can then be used to assign a p-value to the likelihood that the null hypothesis may be rejected. We used this method to assess the distribution of the global network measures for each of the 6 groups: controls ($\times 3$ - aripiprazole, placebo, sulpiride) and patients ($\times 3$ - aripiprazole, placebo, sulpiride). A p-value of $<0.05$ was taken to indicate a significant result.

\subsubsection*{Analysis of variance (ANOVA)}

To examine the effects of the drugs, volunteer type and task difficulty on cognitive performance, we performed a 3 way ANOVA with 2 repeated measures \citep{cohen2007} in various combinations on the hit rates of the subjects. A p-value of $<0.05$ was taken to indicate a significant result.

\subsection*{Anatomical parcellation and wavelet decomposition}

For each individual dataset, voxel time series were averaged within each of the 325 equally sized anatomical regions in a random driven atlas (see \citep{zalesky2010} for approach). 28 regions lacked good quality fMRI timeseries for some subjects so were discarded from our analysis, leaving datasets for 297 brain regions for all subjects. The maximal overlap discrete wavelet transform \citep{percival2000} was used to decompose each individual regional mean fMRI time series into the frequency interval 0.030 - 0.060 Hz (scale 3). This frequency range was selected as it is has been shown that frequencies of $\leq$ 0.1 Hz exhibit the most prominent salient neuronal fMRI dynamics \citep{achard2006}.

\subsection*{Topological network construction}

Undirected weighted networks were generated for each individual based on correlating scale 3 wavelet coefficients. The resulting correlation coefficients $r_{ij}$ form the weight of the edges connecting regions $i$ and $j$. A simple thresholding procedure was then applied to eliminate edges with weights smaller than $ \tau $; all remaining edges are then given a weight of $1$, providing an undirected, unweighted network. The threshold $ \tau$ can be varied to generate networks with any desired percentage of possible connections. Following the example of \citep{lynall2010}, which studies a subset of this particular dataset, we choose to focus on $37\%-50\%$ connectivity. Firstly, this ensures that all graphs are connected, and secondly, it avoids the increasing randomness associated with higher connection densities \citep{humphries2006}. All results given for the unweighted networks are averages across this range.

\subsection*{Global network measures}

\subsubsection*{Clustering}

The \emph{clustering coefficient}, $C_{i}$, for a node $i$ can be defined as \citep{watts1998}:

\begin{equation}
C_{i}=\frac{2T(i)}{k_{i}(k_{i}-1)}
\end{equation}

where $T(i)$ is the number of triangles containing node $i$, and $k_{i}$ is the degree of $i$ - see Figure \ref{fig:fig1}(a).

The \emph{average clustering} then provides a global network measure of clustering, and is simply the average of all values of nodal clustering, or:

\begin{equation}
C=\frac{1}{N}\sum_{i \in G}{C_{i}}
\end{equation}

\subsubsection*{Characteristic path length}

If the shortest path lengths, $L_{ij}$, between all existing node pairs $i$ and $j$, are identified, then the \emph{characteristic path length} of the network, $L$, is simply given by the mean of their sum:

\begin{equation}
L=\frac{1}{N(N-1)}\sum_{i \neq j \in G}{L_{ij}}
\end{equation}

\subsubsection*{Efficiency} 

A measure of the \emph{global efficiency} of a network, $E_{Global}$, is given by the mean of the sum of the inverse shortest path lengths, $L_{ij}$,
between all existing node pairs $i$ and $j$ \citep{achard2007, latora2001}:

\begin{equation}
E_{Global}=\frac{1}{N(N-1)}\sum_{i \neq j \in G}\frac{1}{L_{ij}}
\end{equation}

where $N$ is the number of nodes in the graph $G$. Networks for which the
average path length is small can thus be said to have high global efficiency \citep{achard2007} (see Figure \ref{fig:fig1}(a)). 

This is equivalent to averaging the nodal efficiencies for all nodes in the network.

\subsubsection*{Assortativity}

The \emph{assortativity} of a network is a measure of the preference of its nodes to connect to other nodes of similar degree. Let $e_{xy}$ be the joint probability distribution (or mixing matrix) of the degrees. Then if $\sum_{y} e_{xy} = a_x$ and $\sum_{x} e_{xy} = b_y$ are, respectively, the fraction of edges that start and end at vertices with values $x$ and $y$ and further that $ e_{xy} \neq a_{x}b_{y}$ (the case of no assortative mixing), the assortativity coefficient can be defined simply by calculating the Pearson correlation coefficient \citep{newman2003}:

\begin{equation}
A=\frac{\sum_{xy} xy(e_{xy} - a_{x}b_{y})}{\sigma_{a} \sigma_{b}}
\end{equation}

where $\sigma_{a}$ and $\sigma_{b}$ are the standard deviations of the distributions $a_x$ and $b_y$. $A$ has a value in the range $ -1 \leq r \leq 1 $, where $A=1$ would correspond to a perfect correlation between $x$ and $y$, ie perfect assortativity, and similarly $A=-1$ would indicate perfect disassortativity.

\subsection*{Software}

Motion diagnostics, preprocessing and parcellation of the functional MRI data were completed using the preprocessing pipeline with temporal despiking from \citep{patel2014,kundu2012}. Metric calculations and network manipulations were carried out using the Python networkx library \citep{networkx} and Matlab. We used IBM SPSS Statistics for all ANOVA calculations \citep{spss2012}.

\newpage

\section*{Results}

\begin{figure}
\begin{center}
\includegraphics[width=15.5cm]{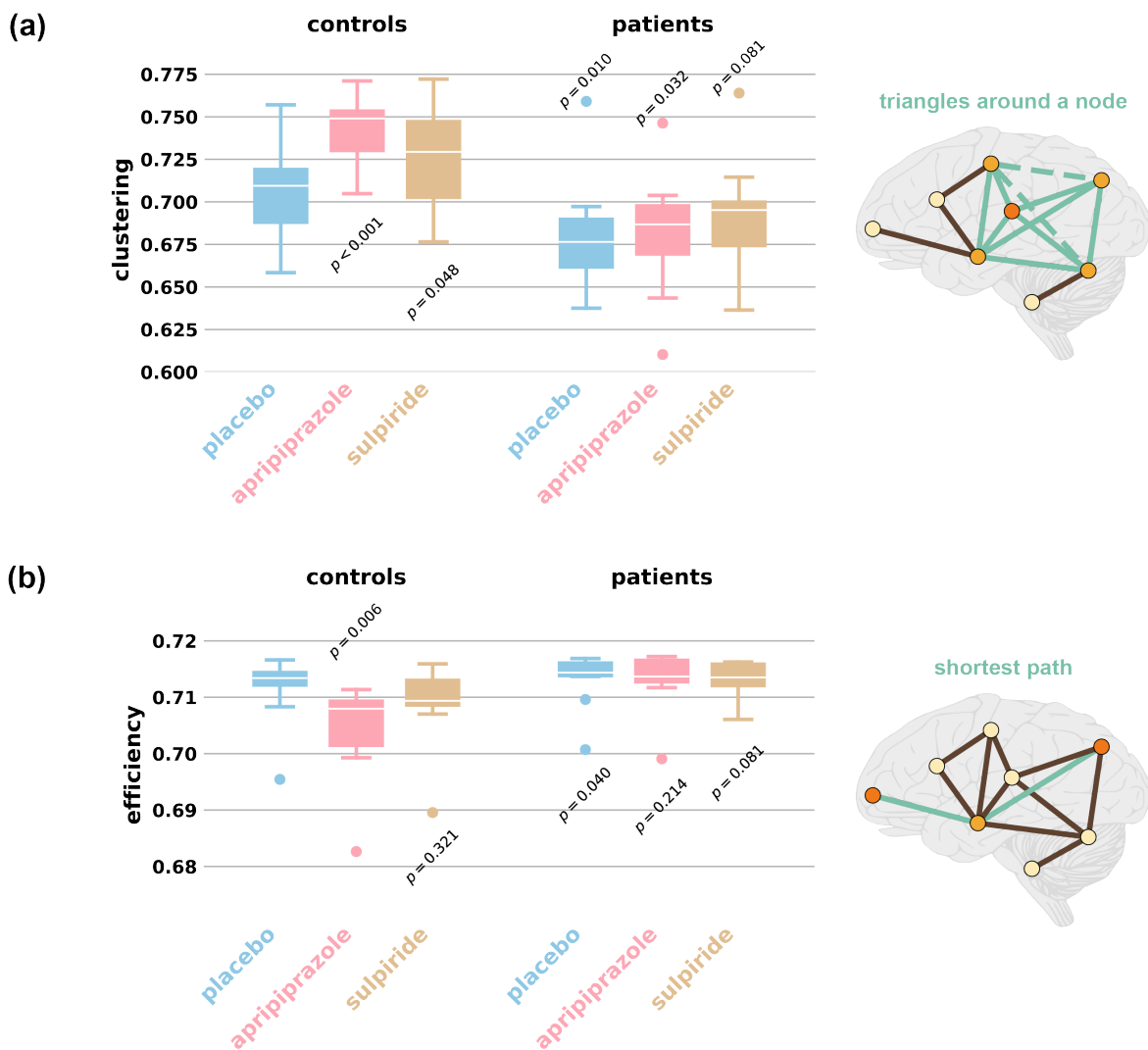}
\end{center}
\caption{{\bf Average clustering and global efficiency values.} The box plots display distributions of (a) average clustering and (b) global efficiency for the networks of each group and drug. Schizophrenia is associated with lower clustering and higher efficiency. The antipsychotic medications increase clustering and decrease efficiency, therefore moving patients closer to controls and affecting the control networks. The schematics illustrate the concepts of (a) clustering and (b) efficiency. (a) Clustering measures the number of triangles which exist around around a node (green solid lines), as a proportion of those that could (also green dashed lines). (b) Efficiency averages the inverse shortest paths (green lines) between all node pairs; many short paths equates to higher efficiencies.}
\label{fig:fig1}
\end{figure}

\subsection*{The effects of schizophrenia on global efficiency and clustering are mitigated by medication}

We first compared the functional brain networks derived from schizophrenia patients and healthy volunteers on placebo treatments, and as expected, \citep{alexanderbloch2013, lynall2010} find distinct differences. In agreement with the existing literature, the schizophrenia networks have increased efficiency (average $E_{Global,SZ}=0.714$ compared to $E_{Global,HV}=0.712$, $p=0.04$) and reduced clustering (average $C_{SZ}=0.674$ compared to $C_{HV}=0.707$, $p=0.01$) -€" see Figure \ref{fig:fig1}.

We next employed an ANOVA (two-way, one repeated measure) to examine any differences in network measures between groups, and the factors underlying them. This confirmed group differences due to subject type on both efficiency and clustering ($p=0.010$ and $p=0.002$ respectively), and also indicated group differences due to drug effect ($p=0.041$ and $p=0.05$ respectively) - see Tables \ref{tab:table1} and \ref{tab:table2}. Our hypothesis was that the antipsychotic medications would aim to make the brain connectivities of patients more similar to those of healthy individuals. In the light of the observed differences between the control and patient placebo groups, this hypothesis translates to an expectation that aripiprazole and sulpiride will reduce efficiency and increase clustering. We do indeed find this, both for patients and for healthy volunteers (Figure \ref{fig:fig1}), with the only exception of when patients were treated with sulpiride, for which there was no effect on global efficiency. Moreover, after each drug treatment the brain networks of people with schizophrenia had global efficiencies which were no longer statistically different from the healthy brain networks (average $E_{Global}=0.713$, $p=0.241$ for aripiprazole and $E_{Global}=0.715$, $p=0.081$ for sulpiride). Both drugs also led to average clustering coefficients that were much closer to those of healthy brain networks (average $C=0.681$, $p=0.032$ for aripiprazole, $C=0.677$, $p=0.081$ for sulpiride).

\begin{table}[h]
\begin{center}
\begin{tabular}{|l|l|l|l|l|l|}
\hline
\textbf{Source}              & \textbf{SS} & \textbf{df} & \textbf{MS} & \textbf{F} & \textbf{p}      \\ \hline
\textbf{Between groups}            &      &          &             &            &                 \\
Subject type    &        0.026     &     1        &      0.026       &       12.208    &      \color{red}{0.002}         \\
Error &         0.040    &      19       &    0.002         &            &                 \\
\textbf{Within groups} &	&	&	&	& \\
Drug                         & 0.004     & 2           & 0.002      & 6.023    & \color{red}{0.005 }      \\
Drug*Subject type                & 0.004     & 2          & 0.002      &       5.997     &   \color{red}{0.005}              \\ 
Error &                    0.012     &     38        &     $<0.001$       &          &       \\ \hline

\end{tabular}
\end{center}
\caption{{\bf Summary statistics for a 2 way ANOVA with 1 repeated measure on the network global clustering values of patients and healthy controls treated with placebo, aripiprazole, and sulpiride.}  Individuals for which networks were available for all drug treatments were used. There is a significant difference between the network clustering of the HV and SZ groups ($p = 0.002$), a significant drug effect ($p = 0.005$) and an additional drug-group type interaction term ($p = 0.005$) - all highlighted in red. This interaction term stems from the effect of aripiprazole - it greatly increases the clustering of control networks whilst causing only a small and variable increase in the schizophrenia networks. The placebo and sulpiride treatments have a more consistent effect on the two groups.}

\label{tab:table1}
\end{table}

\begin{table}[h]
\begin{center}
\begin{tabular}{|l|l|l|l|l|l|}
\hline
\textbf{Source}              & \textbf{SS} & \textbf{df} & \textbf{MS} & \textbf{F} & \textbf{p}      \\ \hline
\textbf{Between groups}            &      &          &             &            &                 \\
Subject type    &        $<0.001$     &     1        &      $<0.001$        &       8.093     &      \color{red}{0.010}           \\
Error &         0.001    &      19       &    $<0.001$         &            &                 \\
\textbf{Within groups} &	&	&	&	&\\
Drug                         & $<0.001$      & 2           & $<0.001$       & 3.480    & \color{red}{0.041}        \\
Drug*Subject type                & $<0.001$      & 2          & 0.0001     &       1.005     &   0.376              \\ 
Error &                    0.0010    &     38        &     $<0.001$        &          &       \\ \hline

\end{tabular}
\end{center}
\caption{{\bf Summary statistics for a 2 way ANOVA with 1 repeated measure on the network global efficiency values of patients and healthy controls treated with placebo, aripiprazole, and sulpiride.} Individuals for which networks were available for all drug treatments were used. There is a significant difference between the network efficiency of the HV and SZ groups ($p = 0.010$) and a significant drug effect ($p = 0.041$) - highlighted in red. }
\label{tab:table2}
\end{table}

\subsection*{Aripiprazole significantly changes healthy brain networks}

We found that aripiprazole has a dramatic effect on healthy individuals, with a large variation across individuals. We observe significantly reduced global efficiencies (average $E_{Global}=0.704$, $ \sigma=0.009 $, $p=0.006$), and a considerable increase in clustering ($C=0.743$, $p \textless 0.001$). Sulpiride increased clustering in healthy networks (average $C=0.726$, $p=0.028$), but had no significant effect on global efficiency (average $E_{Global}=0.709$, $p=0.321$). Almost all metrics examined are greatly altered in the healthy volunteers administered aripiprazole, indicating considerable restructuring of functional connectivity: see Figure \ref{fig:fig1}). These results are consistent with observations in the brain networks of people with schizophrenia: the drug treatments reduce efficiency and increase clustering.

\subsection*{Cognitive performance of healthy individuals is impaired after taking aripiprazole}

\begin{figure}
\begin{center}
\includegraphics[width=10cm]{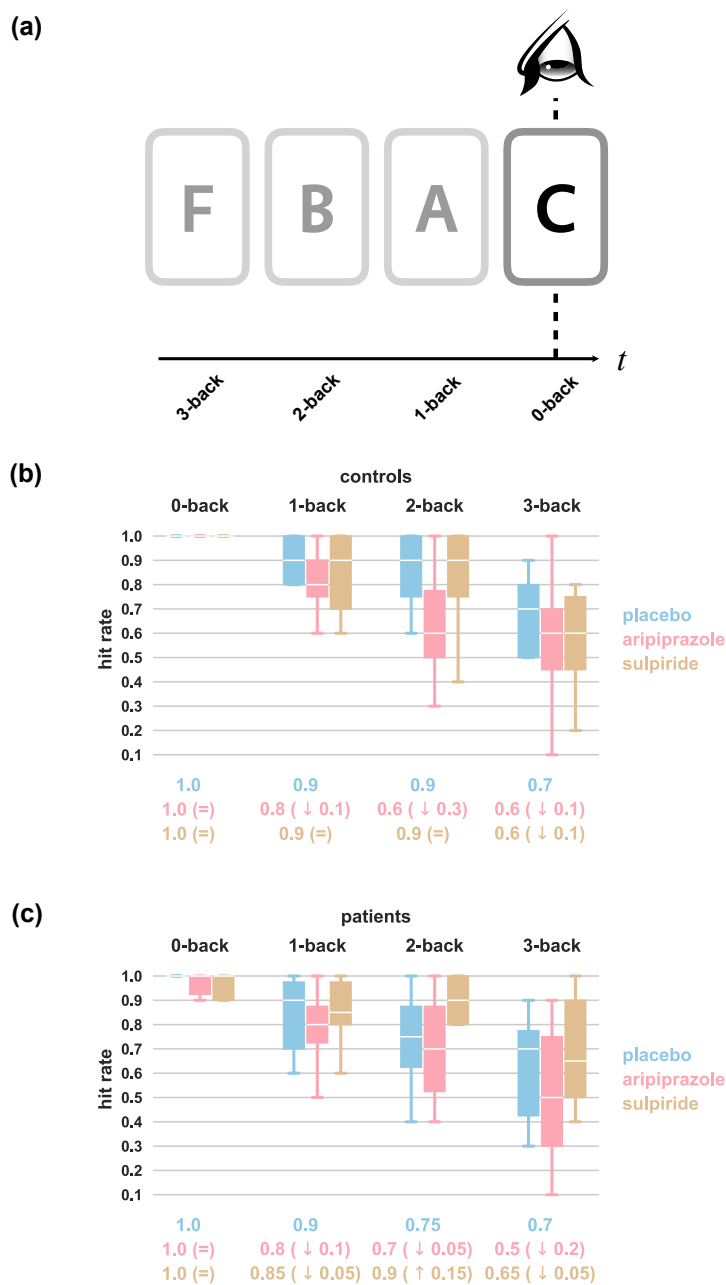}
\end{center}
\caption{{\bf N-back working memory task.} Panel (a) illustrates the nature of the task. Subjects are shown a sequence of letters, and asked to recall the letter which is ``N-back''. In the example shown, the correct answer to the 1-back version is A, to the 2-back version B, etc. For each drug treatment, hit rates from (b) healthy individuals and (c) schizophrenia patients averaged across each level of difficulty are presented. Values below the boxes represent the median values, and for the drug treatment groups, the difference with the median of the placebo group is provided in brackets. Aripiprazole is associated with a reduced number of correct answers as compared to placebo.}
\label{fig:fig2}
\end{figure}

All subjects score consistently highly on the very easy 0-back task, but their performance deteriorates considerably with increasing difficulty of the task, with subjects experiencing profound difficulty with the 3-back version (see Figure \ref{fig:fig2}). In the healthy cohort, aripiprazole has a detrimental effect on performance, whereas the impact of sulpiride is negligible - see Figure \ref{fig:fig2}(b). In the challenging 2-back version of the task, the disparity is most clear - aripiprazole has a negative impact on cognitive performance, giving rise to an average hit rate of $ 0.65 \pm 0.21 $ (compared to $ 0.83 \pm 0.19 $ on placebo). Sulpiride, however, has no noticeable impact, with subjects achieving an average hit rate of $ 0.83 \pm 0.18 $.

An ANOVA (three-way, two repeated measures) revealed that, naturally, the predominant factor in determining success at the working memory tests was the difficulty of the task ($ p \textless 0.0001 $) - see Table \ref{tab:table3}. However, it also demonstrated that the drug treatments have a significant effect ($p=0.007$). We then separated out the drug treatments into placebo-aripiprazole and placebo-sulpiride groups and repeated the ANOVA on controls and patients separately. This shows that the effect is only observed in the healthy volunteers, and that aripiprazole is the medication responsible ($p=0.015$ for the placebo-aripiprazole groups and $p=0.769$ for the placebo-sulpiride groups). Thus, amongst healthy volunteers, we find that aripiprazole results in poorer performance at the N-back working memory task as compared to placebo, and that sulpiride has no noticeable effect.

\begin{table}[h]
\begin{center}
\begin{tabular}{|l|l|l|l|l|l|}
\hline
\textbf{Source}              & \textbf{SS} & \textbf{df} & \textbf{MS} & \textbf{F} & \textbf{p}      \\ \hline
\textbf{Between groups}            &      &          &             &            &                 \\
Subject type    &        0.007     &     1        &      0.007       &       0.044    &      0.836           \\
Error &         3.762    &      23       &    0.164       &            &                 \\
& & & & & \\
\textbf{Within groups} &	&	&	&	& \\
Drug                         & 0.386   & 2           & 0.193     & 5.489   & \color{red}{0.007}     \\
Drug*Subject type                & 0.011    & 2          & 0.006     &       0.160    &   0.853            \\ 
Error &                    1.618     &     46       &     0.035       &          &       \\ 
& & & & & \\
Cog. difficulty                        & 4.828  & 3           & 1.609     & 42.779   & \color{red}{ \textless 0.0001}      \\
Cog. difficulty*Subject type                & 0.045     & 3          & 0.015     &       0.403    &   0.751            \\ 
Error &                    2.596     &     69       &     0.038      &          &       \\ 
& & & & & \\
Drug*Cog. difficulty                         & 0.120   & 6           & 0.020     & 1.160   & 0.331      \\
Subject type*Drug*Cog. difficulty              & 0.162    & 6          & 0.027     &       1.160    &   0.331             \\ 
Error &                    2.384     &     138       &     0.017     &          &       \\  \hline

\end{tabular}
\end{center}
\caption{{\bf Summary statistics for a 3 way ANOVA with 2 repeated measures on the hit rates during a working memory task with 4 levels of difficulty of patients and healthy controls treated with placebo, aripiprazole, and sulpiride.} Individuals for which data were available for all drug treatments were used. We see a significant effect of cognitive difficulty ($p < 0.001$) and a significant drug effect ($p = 0.007$) - highlighted in red.}
\label{tab:table3}
\end{table}

\subsection*{Cognitive tests show worse performance of patients but do not capture drug effects}

Patients scored worse than controls in the N-back working memory task, but not significantly so, with an average hit rate of $0.73 \pm 0.18$ for patients and $ 0.83 \pm 0.19 $ for controls for the placebo 2-back task. Aripiprazole and sulpiride do not change the performance of patients much (average hit rates of $0.71 \pm 0.21$ and $0.79 \pm 0.28$ respectively for the 2-back task) with an ANOVA showing no significant drug effect ($p=0.217$) - see \ref{fig:fig2}(c). 

\newpage

\section*{Discussion} 

\subsection*{Network topology, illness, and medication}

It has been known for some time that the functional brain network organisation of people with schizophrenia differs from that of healthy volunteers \citep{alexanderbloch2013, lynall2010}. However, very few studies have been conducted into the effect of antipsychotic medication, or indeed any drug for any brain disorder, on this network organisation. The ones that have been conducted found measurable drug effects \citep{achard2007,schmidt2013,yang2014,flanagan2018} (including, converse to treatment, inducing psychosis \citep{lahti2001}). We hypothesised that a drug designed to treat schizophrenia would modify the brain connectivities of patients, making them more similar to those of healthy individuals. The result for global efficiency and clustering in Figure \ref{fig:fig1} clearly demonstrate this principle - sulpiride and aripiprazole act to reduce efficiency in both controls and patients. This leaves the patients with network efficiencies and clustering comparable to those of the unmedicated controls, and the controls with lower efficiency and higher clustering than before. In addition we also observed a strong effect of aripiprazole on healthy controls (see Figure \ref{fig:fig1} and Tables \ref{tab:table1} and \ref{tab:table2}. This finding is consistent with our result in patients, as the drug seemingly tries to `correct' for schizophrenia network characteristics - in the absence of schizophrenia - by altering the network metrics to decrease efficiency and increase clustering..

\subsection*{Network topology and cognitive ability}

Our results suggest that there is an optimal configuration for a brain network in terms of maximising cognitive ability: performance worsens given any change (increase {\it or} decrease) in the examined metrics. We saw that as well as having a characteristically different brain network structure, schizophrenia patients perform less well at tests of cognitive ability than their healthy counterparts, as has been previously demonstrated \citep{bullmore2012}. Further, we showed that the group who performed most differently on medication (healthy volunteers having been administered aripiprazole) was also the group who had the most changes to the topology of their brain networks. This supports the notion that one's cognitive ability is intrinsically linked to the structure of the brain's functional network \citep{bassett2009}. For example, to integrate and process information quickly, a network requires some level of efficiency. The control group on aripiprazole had diminished efficiency and performed significantly worse at the N-back tasks than when on placebo. No such impaired performance is seen for sulpiride, for which the reduction in efficiency was negligible. On the other hand, the schizophrenia brain networks appear to have {\it too} high an efficiency, perhaps leading to disordered or overwhelming information integration, and they too perform worse. The drugs do decrease efficiency in patients a little and their performance is slightly improved at the 1- and 2-back tasks. A previous study on MEG derived networks using the N-back working memory paradigm \citep{kitzbichler2011} demonstrated a shift towards a more random network configuration (with a decrease in modularity and clustering, and an increase in global efficiency) as the cognitive demands of the task increased. The authors also note significant differences from purely random networks and argue that global sychronisation is important in higher cognition, which is reflected in the network architecture.

\subsection*{Brain networks as a means to assess medication}

This quite unique dataset allowed for an investigation into the effects of antipsychotic drugs on both the large-scale functional brain networks and the cognitive performance of people diagnosed with chronic schizophrenia and healthy volunteers. Despite its limitations, clear drug effects were observed on the network topology and performance at the N-back working memory task. We find a restoration of the healthy network properties in the patient networks, and that aripiprazole impairs cognitive ability and radically rewires the brain networks of healthy volunteers. It would be highly beneficial for future studies to use state of the art functional MRI data to further investigate the links between disrupted networks in people with brain disorders, how medication influences these, and an ``ideal" network topology for the brain (which should be identified by association with an optimal behavioural parameter, such as the best cognitive performance). There is potential for not just diagnosis of the original brain disorder, but also for the quantification of the effectiveness of a drug in treating the illness, and even guidance on its optimal dosage. This would then allow for a systematic comparison between alternative treatments.

\end{document}